\documentclass[conference]{IEEEtran}
\IEEEoverridecommandlockouts
\usepackage{filecontents}
\usepackage{array}
\usepackage{lipsum}
\usepackage{cite}
\usepackage{amsmath,amssymb,amsfonts}
\usepackage{algorithmic}
\usepackage{dblfloatfix}    % To enable figures at the bottom of page
\usepackage{makecell}%To keep spacing of text in tables
\setcellgapes{4pt}
\usepackage{graphicx}
\usepackage{multirow} 
\usepackage{textcomp}
\usepackage{xcolor}
\usepackage{float}
\usepackage{subfig}
\def\BibTeX{{\rm B\kern-.05em{\sc i\kern-.025em b}\kern-.08em
    T\kern-.1667em\lower.7ex\hbox{E}\kern-.125emX}}
\begin{document}

\title{Pixel-wise Segmentation of Right Ventricle of Heart\\
%{\footnotesize \textsuperscript{*}Note: Sub-titles are not captured in Xplore and should not be used}
\thanks{We thank Nvidia corporation for providing GPU grant for this work.}
}

% \author{\IEEEauthorblockN{Yaman Dang}
% \IEEEauthorblockA{\textit{Indian Institute of Technology Bombay}\\
% Mumbai, India \\
% yamandang@gmail.com}
% \and
% \IEEEauthorblockN{Deepak Anand}
% \IEEEauthorblockA{\textit{Indian Institute of Technology Bombay}\\
% Mumbai, India \\
% deepakanand@iitb.ac.in}
% \and
% \IEEEauthorblockN{Amit Sethi}
% \IEEEauthorblockA{\textit{Indian Institute of Technology Bombay}\\
% Mumbai, India \\
% asethi@iitb.ac.in}
% }

\author{Yaman~Dang{*}\thanks{{*} Joint co-first authors. Email- yamandang@gmail.com and {deepakanand,asethi}@iitb.ac.in}, Deepak~Anand{*}, Amit~Sethi\\
\IEEEauthorblockA{\textit{Electrical Engineering, IIT Bombay}}}
\maketitle

\begin{abstract}
One of the first steps in the diagnosis of most cardiac diseases, such as pulmonary hypertension, coronary heart disease is the segmentation of ventricles from cardiac magnetic resonance (MRI) images. Manual segmentation of the right ventricle requires diligence and time, while its automated segmentation is challenging due to shape variations and ill-defined borders. We propose a deep learning based method for the accurate segmentation of right ventricle, which does not require post-processing and yet it achieves the state-of-the-art performance of 0.86 Dice coefficient and 6.73 mm Hausdorff distance on RVSC-MICCAI 2012 dataset. We use a novel adaptive cost function to counter extreme class-imbalance in the dataset. We present a comprehensive comparative study of loss functions, architectures, and ensembling techniques to build a principled approach for biomedical segmentation tasks.
\end{abstract}


\section{Introduction}
The segmentation of ventricles from cardiac magnetic resonance (MRI) images is an important step in the diagnosis of most cardiac diseases, such as pulmonary hypertension, dysplasia, coronary heart disease, and cardiomyopathies. Right ventricle (RV) imaging is particularly difficult because of its anatomy and the motion of the heart.  The RV segmentation is challenging because (i) fuzziness of the cavity borders due to blood flow and partial volume effect, (ii) the presence of wall irregularities in the cavity, which have the same grey level as the surrounding tissues, (iii) the complex crescent shape of the right ventricle, which varies with the imaging slice level~\cite{petitjean2015right}. The segmentation of the right ventricle is currently performed manually by the cardiologists in heart clinics. It is a tiring task that requires about 15 to 20 minutes by a specialist~\cite{petitjean2015right} and is also prone to inter as well as intra-expert variations. Due to this RV functional assessment has been considered secondary to that of the left ventricle (LV), leaving the RV segmentation problem wide open.

Most of the solutions to RV segmentation problem till date are conventional approaches~\cite{guo2018local,punithakumar2015right,moolan2016right,ringenberg2014fast} that although performed well but have not been able to reach human-level accuracy (0.90 Dice score)~\cite{petitjean2015right}. Most of these techniques are data specific and need a lot of modifications to work on different datasets. The deep learning approaches ~\cite{borodin2018right,winther2018nu,luo2016deep,tran2016fully} have matched the performance of traditional approaches without the need for a lot of changes, but they have a long way to go to surpass the human-level accuracy.  Although in some cases, the results are not as good as traditional approaches, but the deep learning algorithms can generalize better on independent datasets.

In this paper, we propose to solve the RV segmentation problem with a deep learning based approach. However, there are many choices of architecture, losses, and ensembling techniques which present a need to establish a principled way to choose best-performing ones from the available choices. We provide an extensive comparative study which spans various loss functions, network architecture, and ensembling techniques. We found that a plain UNet~\cite{ronneberger2015u} architecture trained with an adaptive loss that we propose performed better than other sophisticated models like GCN~\cite{peng2017large} and dilated UNet~\cite{github_mri}.
 
Rest of the paper is organized in the following way. In section \ref{sec:Prev}, we briefly review the literature of the RV segmentation covering the traditional and deep learning based approaches. In section \ref{sec:material}, we present the details of the dataset, models, loss functions, training strategy used for our work. We present our comparative results across various loss functions, architectures, and ensembling strategies in section \ref{sec:results}. We finally conclude in section \ref{sec:conclusion}.

\section{Previous Works}\label{sec:Prev}

The literature of RV segmentation is less abundant as compared to that of the LV segmentation~\cite{petitjean2011review}. Conventional approaches that have been applied to the LV segmentation include graph cuts, deformable models, and level sets, as well as atlas-based methods. While most of these are suited well to the LV geometry, they are insightful for successful RV segmentation. For RV segmentation grid generation ~\cite{punithakumar2015right} or shape model based~\cite{moolan2016right} are particularly popular. A newly proposed hybrid graph-based technique by ~\cite{oghli2018hybrid} takes per-slice segmentation results and combines them to give overall ventricle segmentation. However, we propose a deep learning based approach for RV segmentation, aiming better generalization and usability.

 One of the most popular deep learning models that work well for a large number of biomedical segmentation problems is UNet~\cite{ronneberger2015u}. The ~\cite{github_mri} explores the UNet model for RV segmentation. They used a baseline UNet network. Also ~\cite{borodin2018right} and~\cite{github_mri} explores the dilated version of UNet network. Reportedly the dilated UNet shows better results than the original UNet due to its larger field of view without an increase in the number of parameters. Modified UNet in ~\cite{winther2018nu}, where a single convolution layer replaces the double convolution layers at each depth matches the performance in ~\cite{github_mri}. A fully convolutional network(FCN) based approach was adopted by ~\cite{tran2016fully}. However, a major drawback in it is that the decoder takes into account only high-level features, and all low-level features are neglected. The ensemble of deep learning models is particularly successful for decision making on the corner cases. Majority voting and average voting are two popular choices of ensembling. Another stream of approaches involves two-stage methods, which we explain next. 

 In ~\cite{luo2016deep}, a method based on cascading a CNN network for the separation of region-of-interest(ROI) and pixel-wise classification for the object masks by another CNN was proposed. Similarly, ~\cite{avendi2017automatic} proposes a CNN and stacked autoencoder based method, which locates the ROI and then delineates the boundary of the RV chamber.

The review of the literature suggests many choices of architectures in the domain of deep learning were applied to for RV segmentation. Various network augmentations, post-processing steps such as fully-connected conditional random field (CRF)~\cite{krahenbuhl2011efficient}, training techniques such as using cyclic learning rate schedulers and an ensemble of various models can improve performance of deep learning models on RV segmentation. We present a thorough comparison of performance by adopting techniques which vary in architecture, losses, post-processing, and their aggregation strategies in ensembles.

\section{Materials and Methods}\label{sec:material}

 We adopted a U-Net~\cite{ronneberger2015u} architecture as our baseline approach for RV segmentation. We experiment with various modifications to a baseline model such as boundary refinement layers, loss function, training schedules, and ensemble strategies to guide the selection of the best combination in principle way.

\subsection{Dataset}
 The Right Ventricle Segmentation Challenge (RVSC) was hosted in March 2012. It finally ended up with the ‘‘3D Cardiovascular Imaging: a MICCAI segmentation challenge’’ workshop that was organized in conjunction with the $15^{th}$ edition of MICCAI, held on October $1^{st}$, 2012. The dataset was collected from June 2008 to August 2008. All patients were above the age of 18 and were free from any cardiac conditions. The total of 48 patients participated in this study. The average age ($\pm$ standard deviation) of the participants was $52.1 (\pm 18.1$) years. The cardiac MR examinations were performed with repeated breath-holds for a period of 10-15 s. All MR images were taken with a total of 10-14 contiguous cine short axis slices from the base to the top of the ventricles. The images are of sizes $256 \times 216$ pixels, and further 20 images were taken per cardiac cycle.

\begin{figure}[tbh!]
    \hspace{-1cm}
    \centering
    \includegraphics[scale=0.6]{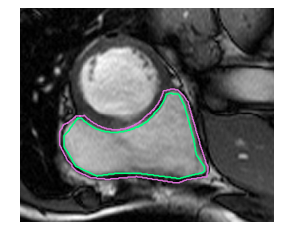}
    \caption{Typical image from dataset with endocardial and epicardial walls~\cite{github_mri}. The endocardial contour is shown in green colour and the epicardial in pink colour (Best viewed in colour).}
    \label{fig:fig3_2}
\end{figure}

 The 48 patients were split equally into three parts such that one part is for training and the other two for testing by the organizers. Each patient's MRI includes between 200 and 280 images, with 20 images per cardiac cycle. A sample of the available dataset is as shown in Figure~\ref{fig:fig3_2}. It shows both endocardial and epicardial contours marked on it in green and pink colors respectively. 

 The dataset consists of images from three spatial locations, i.e., basal, apical, and midway. Basal images are the top of the ventricles and contain the widest portion while the apical slice is the last slice with a detectable ventricular cavity. Usually, in case of apical slices, the size of the ventricle is tiny and, it is not easily distinguishable from tissue linings.

\begin{figure}[!th]
    %\hspace{-1cm}
    \centering
    \includegraphics[scale=0.3]{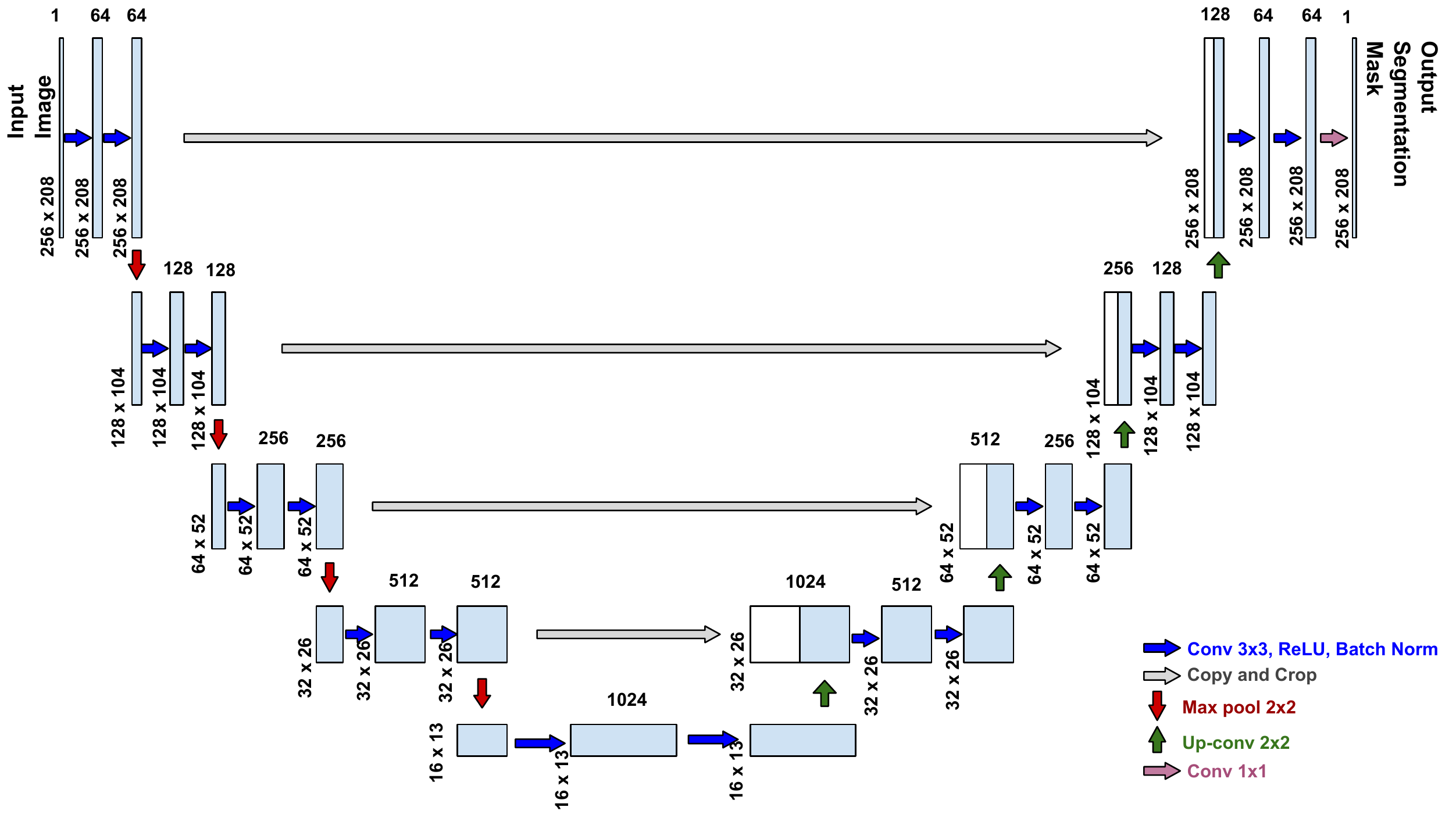}
    \caption{UNet~\cite{ronneberger2015u} architecture used as baseline model.}
    \label{fig:fig5_2}
\end{figure}

\subsection{Data Augmentation}
 The training set consisted of MRI images from 16 patients. We have split the training set of 16 patients in smaller datasets of 12 patients and a validation set of 4 patients. Finally, a total of 181 images are used for training, and 62 for the validation of results. We used data augmentation techniques to increase the number of different types of images for training and reduce overfitting. We adopt horizontal and vertical flips as well as a random rotation in the range of $-180^{o} \text{ to }+180^{o} $. The same transforms were done for pairs of ground-truth masks and input images. We adopted Contrast Limited Adaptive Histogram Equalization (CLAHE) to adapt to the contrast variation across patients. Oversampling of the training images with less than 1\% area to be segmented was also adopted for robust performance.

\subsection{Network Architectures}
In this section, we present the details of the generic choices of network architectures.
\subsubsection{UNet}
The UNet~\cite{ronneberger2015u} architecture is best known to provide excellent segmentation results on various medical imaging datasets. It is a fully-convolutional network with skip connections across different level of depths of encoder and decoder. The architecture that we used has a depth of four. The size of all the kernels used in convolutional layers is $3\times3$, and the upsampling operation is done using bilinear interpolation.

\subsubsection{Dilated UNet}
 As biomedical images consist of specific placements of organs, the global context knowledge of the position of organs helps in better segmentation. In UNet, the field-of-view of the convolutional kernel at the highest depth is also not sufficient to cover the entire image. Thus to increase the field-of-view, we introduce dilation in the convolutional kernels. Dilated UNet is successful in other segmentation tasks~\cite{borodin2018right, github_mri}. The dilation scheme that we used for the UNet is 2,2,2, and 4 starting at the input layer and going to the maximum depth.

\subsubsection{Global Convolution Network (GCN)}
\begin{figure}[tbh!]
\centering
\subfloat[Global Convolution Block ]{\includegraphics[scale=0.3]{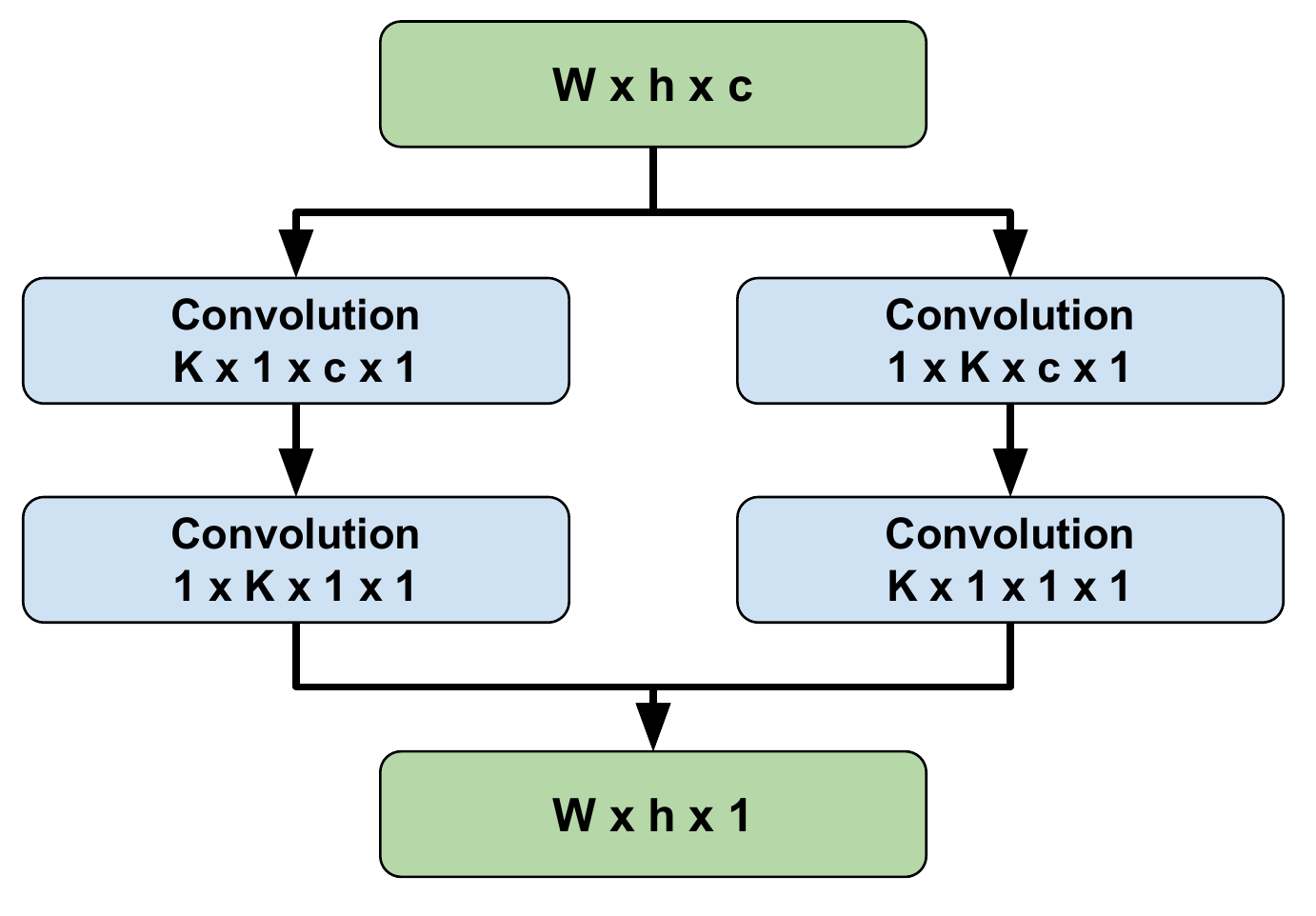}} \hspace{0.5cm}
\subfloat[Boundary Refinement Block ]{\includegraphics[scale=0.3]{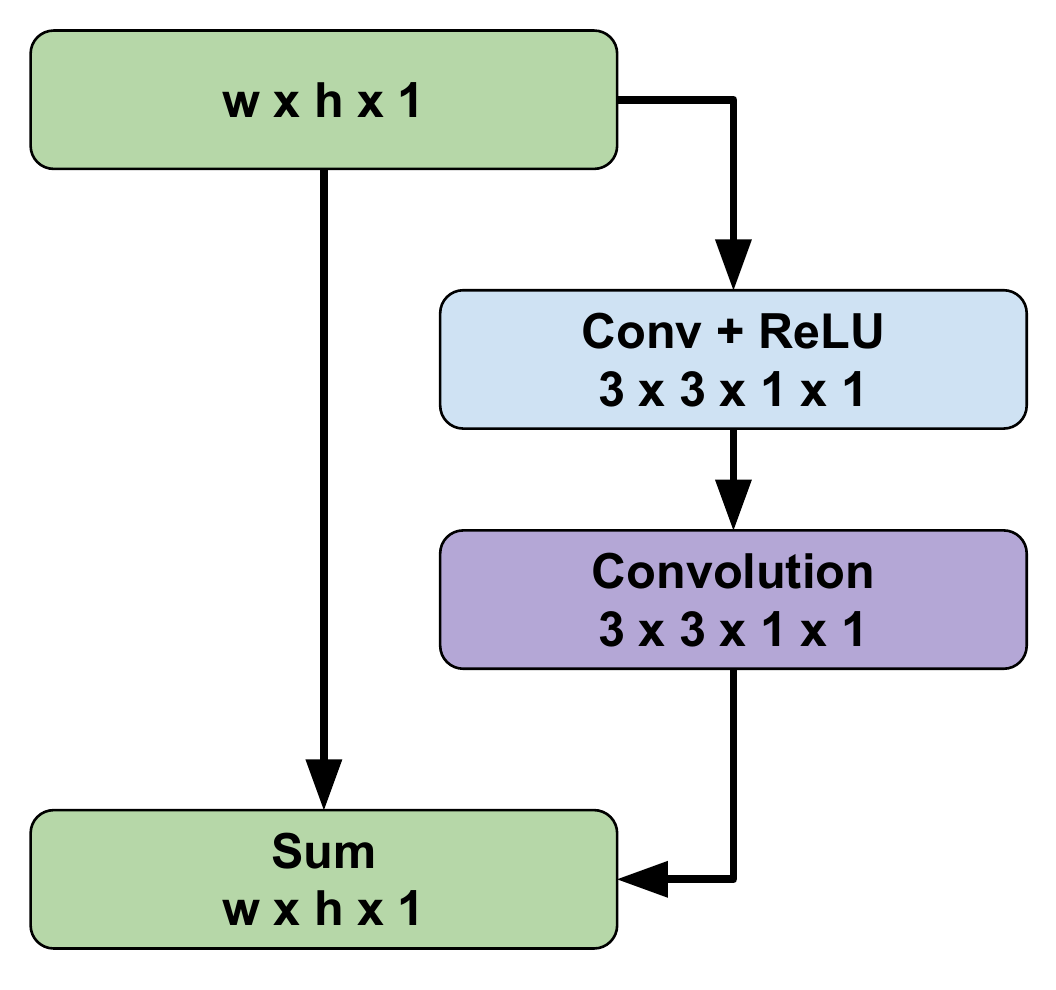}} \hspace{0.5cm} \\
\caption{Structure of GCN and the BR block}
\label{fig:fig6_2_3}
\end{figure}

 The motivation behind using this architecture is that it considers a global view while deciding. To improve localization, ~\cite{peng2017large} proposed GCN Block and the Boundary refinement (BR) block. In GCN block, instead of using a square kernel, it has a combination of $1 \times k + k \times 1$ and $k \times 1 + 1 \times k$ convolutions. This enables dense connections in a large region of k x k without increasing the number of parameters to be learned. The structure of the BR block consists of two $3 \times 3$ convolutional layers without the activation layer. The expression for the final prediction map is $\bar{P} = P + R(P)$ where $P$ is the initial score, and $R(.)$ is the residual layer. The structure of the GCN block and BR block is as shown in Figure \ref{fig:fig6_2_3}.

 The overall structure of the model uses a ResNet as feature extraction network and the fully connected layers as the segmentation networks. We have done various experiments with the size of kernels.

\subsubsection{GCN with UNet feature map}
 This model has a structure very similar to the GCN model. The main difference is that the ResNet based feature extraction block is replaced by the baseline UNet based feature extraction block. The rest of the flow of the GCN and BR layers remain the same.

\subsubsection{UNet with GCN at last layer}
 Seeking inspiration from ~\cite{liu2018semi}, we applied the GCN+BR layer at the end of our baseline UNet model after removing the end convolution layer. We tried experimenting with three different sizes of the kernels, i.e., $9 \times 9$, $5 \times 5$, and $15 \times 15$. This configuration facilitates larger aggregation field-of-view for pixel-wise decision making.

\subsubsection{Double the features in UNet and Dilated UNet}
 To improve our baseline UNet and dilated UNet performance, we tried to increase the number of features generated at each layer. So we doubled the number of kernels used. This considerably increased the size of the model and hence slowed down the training.

\subsection{Losses} The RV segmentation problem is an imbalanced class segmentation problem due to the varying volume of the RV in the MRI images. To account for this imbalance, we adopted four losses and benchmarked their performance against the commonly used Dice loss.
\subsubsection{Focal Loss}
 The focal loss is designed to address scenarios where there is a high imbalance between foreground and background classes (e.g., 1:1000)~\cite{lin2017focal}. It gives an exponentially small weight to well-classified pixels, and weight for the misclassified pixels is almost equal to 1. A modulation factor of $ (1-p_t)^\gamma $ is added to the binary cross entropy loss.
\[ p_t =  \begin{cases}
    p       & \quad \text{if } y = 1 \\
    1-p  & \quad \text{otherwise } 
  \end{cases}  \]
%\[ \text{Cross\, Entropy \,} L_C(p_t) = -log(p_t)   \]
  \[ L_F(p_t) = -(1-p_t)^\gamma \; log(p_t)  = -(1-p_t)^\gamma L_C  \]

 For $ \gamma =0 $, focal loss is equal to cross-entropy loss. We experimented with different values of $\gamma$ i.e. $\gamma =  0.5, 1, \text{and }2 $. Our model performed best for $\gamma = 1 $.

\subsubsection{Cross Entropy + Dice Loss}
 The BCE loss tries to match the background and foreground pixels in the prediction to the ground truth masks, which results in erroneous salt and pepper pattern, i.e., holes in the intended foreground region. BCE loss does not emphasize on keeping the object together while Dice loss considers the entire object by computing the overlap between predicted and ground truth objects. Thus combining both BCE loss and the Dice loss will give us the advantages of both. The disadvantage of Dice loss is that it considers only the foreground pixels while calculating the overlap. We have given equal weight to both these losses.

\subsubsection{Cross Entropy + Dice + Inverse Dice Loss}
 To remove the disadvantage of Dice+BCE combination, we compute the Dice score on the background region. This computation of Dice score only on the background is called as Inverted Dice score. It is the exact opposite of the Dice score, which makes sure that the background region is well classified. We have used an equal combination of all three losses. The following equations gives the expressions for the three losses, i.e. BCE ($L_C$), Dice loss ($L_D$), and inverted Dice loss ($L_I$):

\[ L_{C} = -\sum_{i} \sum_{j} log(p(i,j)) \]
\[  L_{D}=1- \frac{2\sum_{i} \sum_{j} p(i,j)\  g(i,j) + \epsilon}{\sum_{i} \sum_{j} p(i,j) + \sum_{i} \sum_{j} g(i,j) + \epsilon} \]
\[  L_{I}= 1- \frac{2\sum_{i} \sum_{j}(1- p(i,j))\ (1-g(i,j)) + \epsilon}{\sum_{i} \sum_{j} (1-p(i,j)) + \sum_{i} \sum_{j} (1-g(i,j)) + \epsilon} \]
 where $p(i,j)$ is the predicted probability that the pixel at location $(i,j)$ belongs to the foreground and $ g(i,j)$ is the ground truth mask.

\subsubsection{Switching Loss}
 This loss is an optimized combination of the three losses. It is the sum of BCE loss and an adaptive combination of Dice loss $L_D$ and inverted Dice loss $L_I$ as shown in the equation given below. We switch the emphasis to be given to the loss on the basis of the ratio of the number of foreground pixels $C_f$ and the total number of pixels $C_t$.
\[ L_{switching} = \begin{cases}
L_{C} +\lambda \  L_{D} + (1-\lambda)\  L_{I},& \text{ for } \frac{C_n}{C_t}>\tau \\
L_{C} +(1-\lambda) \  L_{D} + \lambda\  L_{I},& \text{ for } \frac{C_n}{C_t}<\tau
\end{cases}
\]
 wherein the hyper parameters have the following ranges: $ {0 \leq \lambda \leq 1} \text{ and }0 \leq \tau \leq 1 $. Experimentally, we found that $\lambda = 0.75$ works best for us. 
 
\subsection{Post-processing Techniques}
Although our final proposed model is post-processing free, we tried fully-connected CRF as a post-processing step on the predicted segmentation maps to provide a  comparative analysis. We also experiment with two ensemble strategies-- majority voting and average voting.

\subsection{Network Training Details}
We use PyTorch~\cite{paszke2017automatic} library for the implementation of the deep learning models. We use the same protocol for all training unless stated otherwise. We used a train to test split ratio of 3:1. For the training, we use Adam optimizer with an initial learning rate of $ 10^{-3} $. The model is trained for 600 epochs. Our training algorithm follows a cyclic learning rate schedule~\cite{loshchilov2016sgdr}, i.e., the value of the learning rate keeps on following a cosine curve between $10^{-3} $ to $10^{-6}$. To reduce the effect of overfitting used L2-regularization with lambda equal to $10^{-3}$. For training loss, we experimented with various combinations, as mentioned in the previous subsection.

\subsection{Evaluation Measures}
Let $p$ and $t$ be the predicted and true contours of the ventricle. Let $P$ and $T$ be the corresponding areas enclosed by contours p and t, respectively. We use the following two evaluation metrics to check the accuracy of the segmentation method.

\subsubsection*{Dice Coefficient} 
The Dice's coefficient is a measure used to compare the similarity between the two binary images. The Dice index normalizes the number of true positive elements to the average size of the sets $P$ and $T$. The Dice coefficient is calculated as follows: 
    
\begin{equation}
    Dice (P,T) =\frac{2|P * T | }{(|P| + |T|)}
\end{equation}

where $*$ is the element-wise multiplication and $|.|$ denotes the number of pixel elements that are '1'. The Dice score value is between 0 and 1. It is 0 for disjoint areas, and 1 for perfect agreement.

\subsubsection*{Hausdorff Distance (HD)}
The Hausdorff distance is a measure of the closeness of two subsets in a metric space. Two sets are close in the Hausdorff distance if every point of either set is close to some point of the other set. The Hausdorff distance is the largest distance from a point in one set to the closest point in the other set. Hausdorff distance can be calculated as: 
\begin{equation}
    d_H(P, T) = max\{ \sup_{p\in P} \inf_{t\in T} d(p,t), \sup_{t\in T} \inf_{p\in P} d(p,t) \}
\end{equation}
where $\sup$ and $\inf$ represents the supremum and infimum respectively and $d(.)$ is the distance.

\section{Experiments and Results}\label{sec:results}
We perform three classes of experiments: a) compare across various loss function keeping the baseline UNet architecture, b) compare across various architecture modifications keeping a fixed loss function, and c) compare our best model with state-of-the-art performing approaches in the literature.

\subsection{Comparison of across various loss functions}
 We experimented with four loss functions keeping the UNet as constant architecture. Our comparative results is compiled in the Table \ref{table:tab7_2_2}. From the table, one can observe that the Focal loss does not perform well on this dataset despite being specifically designed for handling imbalance class problems. The balanced combination of the three losses (BCE, Dice, and inverted Dice) delivers good results, but the customization of the weights of the combination provides a little better results. Thus Table \ref{table:tab7_2_2} concludes the better suitability of the switching loss for the imbalanced class problems. The last three rows of the Table \ref{table:tab7_2_2} present comparison across ensembling strategies, i.e., majority and average voting and post-processing by CRF. We establish that the CRF is not needed. It can be due to the compactness of the RV in appearance. Both ensemble technique gives similar results.

\begin{table}[thh!]
 %\makegapedcells
 \caption{Comparison of average validation Dice scores for both endocardium and epicardium for all loss function. The last three rows compare across the effect of the ensembling strategy and post-processing on the performance. Switching loss achieves the best performance. Note that these results are based on the train-validation split of 12:4 ratio.}
    \centering
    \begin{tabular}{|m{3.5cm}|m{2cm}|m{2cm}|}
 \hline
   \textbf{Loss Function Used} & \textbf{Endocardium (Inner wall) Avg. Dice Score} & \textbf{Epicardium (Outer wall) Avg. Dice Score}\\ [0.5ex] 
 \hline
  Switching loss ($\lambda = 0.75$) &\textbf{86.68 (10.13)} & \textbf{90.34 (5.68)} \\[2ex]
 \hline
 \v BCE + Dice + Inverted Dice &  86.27 (12.07) & 89.17 (10.16) \\ [2ex]
 \hline
 BCE + Dice & 85.24 (12.31) & 89.20 (7.83) \\ [2ex]
 \hline
 Focal loss ($\gamma = 1$) & 82.43 (17.86) & 87.93 (12.40) \\ [2ex]
 \hline
 Ensemble using majority voting(without CRF) & \textbf{87.24 (11.11)} &\textbf{90.43 (7.71)} \\ [2ex]
 \hline
 Ensemble using majority voting(with CRF) & 87.00 (11.35) & 90.53 (6.91) \\ [2ex]
 \hline
 Ensemble using average probability & \textbf{87.32 (11.13)} & \textbf{90.62 (7.12)}\\ [2ex]
 \hline
 \end{tabular}
%  \caption{Comparison of average validation Dice scores for both endocardium and epicardium for all loss function. The last three rows compares across the effect of the ensembling strategy and post-processing on the performance. Switching loss achieves the best performance. Note that these results are based on the train-validation split of 12:4 ratio.}
 \label{table:tab7_2_2}
\end{table}

\subsection{Comparison of Various Architectures}
 In the previous subsection, we concluded that the switching loss gives the best results with the baseline UNet model. Now, to choose a network architecture in a principled way, we experiment with the available choices of network architectures keeping the switching loss fixed. Table~\ref{table:tab7_3} summarizes the comparative results. Basic UNet model trained with a cyclic learning rate achieves the best performance, which reinforces the idea of having a lesser number of parameters as an advantage.
 
\begin{table}
\caption{Comparison table showing average validation Dice scores for Endocardium for all variations of models. The bottom 5 models were trained with cyclic LR for fair comparison. These results are based on train-validation split into 12:4 ratio.}
\begin{tabular}{|c|c|c|}
\hline
\textbf{Model}                    & \textbf{Variation}    & \textbf{\begin{tabular}[c]{@{}c@{}}Endocardium \\ (Avg. Dice score)\end{tabular}} \\ \hline
\multirow{2}{*}{UNet}             & Step LR               & 84.8603                                                                           \\ \cline{2-3} 
                                  & Cyclic LR             & \textbf{85.9557}                                                                           \\ \hline
Dilated UNet                      & Cyclic LR             & 84.8183                                                                           \\ \hline
\multirow{2}{*}{GCN-ResNet}       & Kernel $7 \times 7$   & 84.0168                                                                           \\ \cline{2-3} 
                                  & Kernel $15 \times 15$ & 81.7004                                                                           \\ \hline
\multirow{2}{*}{GCN-UNet}         & Kernel $7 \times 7$   & 81.8659                                                                           \\ \cline{2-3} 
                                  & Kernel $15 \times 15$ & 83.2252                                                                           \\ \hline
\multirow{3}{*}{UNet with GCN-BR} & Kernel $5 \times 5$   & 83.6046                                                                           \\ \cline{2-3} 
                                  & Kernel $9 \times 9$   & 84.8266                                                                           \\ \cline{2-3} 
                                  & Kernel $15 \times 15$ & 79.5601                                                                           \\ \hline
UNet                              & Double kernel count   & 84.5637                                                                           \\ \hline
Dilated UNet                      & Double kernel count   & 83.1449                                                                           \\ \hline
\end{tabular}
% \caption{Comparison table showing average validation Dice scores for Endocardium for all variations of models. The bottom 5 models were trained with cyclic LR for fair comparison. These results are based on train-validation split into 12:4 ratio.}
 \label{table:tab7_3}
\end{table}

\subsection{Comparison with state-of-the-art}
For comparison with the state-of-the-art results, we split our observations based on the datasets into consideration. As in the literature, results are provided on the test as well as the train set (validation), so we present our results similarly. 

\subsubsection{Comparison with train set (validation)}
 The Table \ref{table:tab7_4_1} shows that our ensemble with majority voting model beats the previous state-of-the-art in both performance metrics.  

\begin{table}
\caption{Comparison of Dice Scores with state-of-the-art methods for train (validation) set. We achieved the best performance by majority voting ensembling.}
 \centering
 \begin{tabular}{|m{2.25cm} |m{1cm} |m{1cm} |m{1cm} | m{1cm}|} 
 \hline
\textbf{Method} & \multicolumn{2}{c|}{\textbf{Endocardium}}   & \multicolumn{2}{c|}{\textbf{Epicardium}} \\
\hline
 & Val. Avg. Dice & Val. Avg. HD (in mm) & Val. Avg. Dice & Val. Avg. HD (in mm)\\ 
\hline
Gregory Borodin, et.al. ~\cite{borodin2018right}  & 85.00 & & 83.00 &\\
\hline
H B Winther, et.al. ~\cite{winther2018nu} & 85.00 (7.00) & & 86.00 (6.00) &\\
 \hline
Gongning Luo, et.al. ~\cite{luo2016deep}  & 86.00 (9.00) & 6.9 (2.6)  84.00 (13.00) & 8.9 (5.7) & \\ 
\hline
Majority Voting (without CRF) ours & \textbf{86.93} &\textbf{ 6.5569} & \textbf{90.18} & \textbf{6.2373}\\
 \hline
 Majority voting (with CRF) ours & 86.24 & 6.9621 & 89.91 & 6.5640\\
 \hline
 Avg. probability based ours & 86.62 & 6.6463 & 90.14 &6.3535\\
 \hline
 \end{tabular}
%  \caption{Comparison of Dice Scores with state-of-the-art methods for train (validation) set. We achieved the best performance by majority voting ensembling.}
 \label{table:tab7_4_1}
\end{table}

\subsubsection{Comparison with test sets}
 The Tables \ref{table:tab7_4_2} and \ref{table:tab7_4_3} gives the results for both inner as well as outer wall for both the test sets. The table is split into three sections, first and the second section gives state-of-the-art results from literature using traditional (marked by \textbf{T}) and deep learning (marked by \textbf{D}) based approaches respectively. From the table, we can observe that our results on the test set 1 are not the best but for test set 2 we can get state-of-the-art results and these are in agreement with the results that we obtain on the validation set. %Also, some failure cases for test set 1 have been discussed in the next sections.
\begin{table*}
\caption{Comparison of Dice scores for Endocardium with state-of-the-art methods for TestSet1 and TestSet2. \textbf{T} represents traditional methods, \textbf{D} represents deep learning based methods, ED stands for End diastolic and ES stands for End systolic cycle. We achieve the best performance using majority voting on Test Set 2.}
 \centering
 \begin{tabular}{|m{2.3cm} |m{0.25cm} |m{0.65cm} |m{0.65cm} |m{0.65cm} |m{0.65cm}|m{0.65cm}|m{0.65cm}||m{0.65cm} |m{0.65cm} |m{0.65cm}|m{0.65cm}|m{0.65cm}| m{0.65cm}|}
 \hline
 \multicolumn{14}{|c|}{\textbf{Endocardium}}    \\
 \hline
 & & \multicolumn{6}{c||}{\textbf{Test Set 1}} & \multicolumn{6}{c|}{\textbf{Test Set 2}} \\ 
 \hline
 \textbf{Method} & \textbf{T/D} & \multicolumn{2}{c|}{Overall} & \multicolumn{2}{c|}{ED} & \multicolumn{2}{c||}{ES} & \multicolumn{2}{c|}{Overall} & \multicolumn{2}{c|}{ED} & \multicolumn{2}{c|}{ES} \\ 
\hline
& & Dice & HD & Dice & HD & Dice & HD & Dice & HD & Dice & HD & Dice & HD\\  \hline
Zengzhi Guo, et al ~\cite{guo2018local} & T & 0.85 (0.11) & 8.63 (6.20) & 0.88 (0.08) & 8.03 (5.57) & 0.81 (0.14) & 9.46 (6.93) & 0.87 (0.08) & 6.93 (4.33) & 0.88 (0.08) & 6.94 (5.05) & 0.85 (0.08) & 6.92 (2.94) \\ 
\hline
K Punithakumar, et al ~\cite{punithakumar2015right} & T & 0.83 (0.13) & 7.72 (3.97) & & & 0.77 (0.16) & 9.64 (4.15) & 0.85 (0.15) & 6.49 (4.44) & & & & \\
\hline
J Ringenberg, et al ~\cite{ringenberg2014fast} & T & 0.83 (0.16) & 9.05 (6.98) & 0.88 (0.11) & 7.69 (6.03) & 0.77 (0.18) &  10.71 (7.69) & 0.83 (0.18) & 8.73 (7.62)  & & & &\\
\hline
\hline
Chuck-Hou Yee ~\cite{github_mri} & D & 0.84 (0.21) & & & & & & 0.84 (0.21) & & & & &\\
\hline
M R. Avendi, et al ~\cite{avendi2017automatic} & D &  & & 0.86 (0.11) &  7.80 (4.26) & 0.79 (0.16) & 7.51 (3.66) &  & & 0.86 (0.10) & 7.85 (4.56) & 0.76 (0.20) & 8.27 (4.23)\\
 \hline
Booz, et al ~\cite{tran2016fully} & D & \textbf{0.84 (0.21)} & 8.86 (11.27) & & & & & 0.84 (0.21) & 8.86 (11.27) & & & &\\
 \hline
  \hline
Majority voting (without CRF) ours & D & 0.8170 (0.22) & 10.56 (11.53) & 0.8578 (0.18) & 10.12 (10.78) & 0.7671 (0.25) & 11.09 (12.40) & \textbf{0.8652} (0.19) & \textbf{6.73} (7.42) & \textbf{0.9100 (0.10)} & \textbf{5.82 (4.95)} & \textbf{0.8093 (0.25)} & \textbf{7.88 (9.56)} \\

\hline
Majority voting (with CRF) ours & D & 0.8131 (0.22) & 10.91 (12.05)  & 0.8544 (0.17)  & 10.94 (12.20) & 0.7629 (0.25) & 10.87 (11.92) & 0.8581 (0.19) & 7.65 (9.50) & 0.9008 (0.10) & 7.00 (8.17) & 0.8047 (0.24) & 8.46 (10.92)\\
\hline
Avg probability based ours & D & \textbf{0.8184 (0.21)} & 10.56 (11.21) & 0.8586 (0.17) & 10.34 (10.64) & 0.7693 (0.25) & 10.82 (11.91) & 0.8624 (0.19) & 7.02 (7.50) & 0.9058 (0.10) & 6.28 (5.14) & 0.8081 (0.24) & 7.95 (9.62) \\
\hline
 \end{tabular}
%  \caption{Comparison of Dice scores for Endocardium with state-of-the-art methods for TestSet1 and TestSet2. \textbf{T} represents traditional methods, \textbf{D} represents deep learning based methods, ED stands for End diastolic and ES stands for End systolic cycle. We achieve the best performance using majority voting on Test Set 2.}
 \label{table:tab7_4_2}
\end{table*}

\begin{table*}
 \caption{Comparison of Dice scores for Endocardium with state of the art methods for TestSet1 and TestSet2. \textbf{T} represents traditional methods, \textbf{D} represents deep learning based methods, ED stands for End diastolic and ES stands for End systolic cycle. We achieve the best performance using majority voting on Test Set 2.}
 \centering
 \begin{tabular}{|m{2.3cm} |m{0.25cm} |m{0.65cm} |m{0.65cm} |m{0.65cm} |m{0.65cm}|m{0.65cm}|m{0.65cm}||m{0.65cm} |m{0.65cm} |m{0.65cm}|m{0.65cm}|m{0.65cm}| m{0.65cm}|}
 \hline
 \multicolumn{14}{|c|}{\textbf{Epicardium}}    \\
 \hline
 & & \multicolumn{6}{c||}{\textbf{Test Set 1}} & \multicolumn{6}{c|}{\textbf{Test Set 2}} \\ 
 \hline
 \textbf{Method} & \textbf{T/D} & \multicolumn{2}{c|}{Overall} & \multicolumn{2}{c|}{ED} & \multicolumn{2}{c||}{ES} & \multicolumn{2}{c|}{Overall} & \multicolumn{2}{c|}{ED} & \multicolumn{2}{c|}{ES} \\ 
\hline
& & Dice & HD & Dice & HD & Dice & HD & Dice & HD & Dice & HD & Dice & HD\\  \hline
Zengzhi Guo, et al ~\cite{guo2018local} & T & & & & & & & & & & & & \\ 
\hline
K Punithakumar, et al ~\cite{punithakumar2015right} & T & 0.87 (0.08) & 8.08 (3.80) & & & 0.82 (0.10) & 9.99 (3.85) & 0.88 (0.10) & 6.95 (3.98) & & & &\\
\hline
J Ringenberg, et al ~\cite{ringenberg2014fast} & T & 0.86 (0.11) & 9.60 (7.01) & 0.90 (0.08) & 8.02 (5.96) & 0.82 (0.13) & 11.52 (7.70) & 0.86 (0.14) & 9.00 (7.46) & & & &\\
\hline
\hline
Chuck-Hou Yee ~\cite{github_mri} & D & \textbf{0.88} (0.18) & & & & & & \textbf{0.88} (0.18) & & & & &\\
\hline
MR. Avendi, et al ~\cite{avendi2017automatic} & D &  & & & & &  &  & & & & & \\
 \hline
Booz, et al ~\cite{tran2016fully} & D & 0.86 (0.20) & 9.33 (10.79) & & & & & 0.86 (0.20) & 9.33 (10.79) & & & &\\
  \hline
  \hline
Majority Voting (without CRF) ours & D & 0.8389 (0.20) & 11.86 (13.27) & 0.8648 (0.18) & 11.57 (13.35) & 0.8073 (0.23) & 12.22 (13.21)& \textbf{0.8936 (0.13)} & \textbf{7.11 (6.04)} & \textbf{0.9164 (0.10)} & \textbf{6.50 (5.72)} & \textbf{0.8652 (0.15)} & \textbf{7.87 (6.37)} \\ 
\hline
Majority Voting (with CRF) ours & D & \textbf{0.8430} (0.19) & 11.36 (11.06) & 0.8747 (0.13) & 10.72 (10.38) & 0.8042 (0.23) & 12.14 (11.82) & 0.8873 (0.14) & 7.36 (5.79) & 0.9075 (0.12) & 6.97 (5.49) & 0.8620 (0.16) & 7.85 (6.13)\\
\hline
Avg probability based ours & D & 0.8387 (0.20) & 11.86 (13.08) & 0.8621 (0.18) & 11.72 (13.42) & 0.8100 (0.22) & 12.04 (12.70) & 0.8882 (0.15) & 7.22 (5.98) & 0.9107 (0.12) & 6.67 (5.38) & 0.8602 (0.17) & 7.90 (6.62)\\ \hline
 \end{tabular}
%  \caption{Comparison of Dice scores for Endocardium with state of the art methods for TestSet1 and TestSet2. \textbf{T} represents traditional methods, \textbf{D} represents deep learning based methods, ED stands for End diastolic and ES stands for End systolic cycle. We achieve the best performance using majority voting on Test Set 2.}
 \label{table:tab7_4_3}
\end{table*}
\vspace{-12pt}

\section{Conclusion}\label{sec:conclusion}
We presented a comprehensive approach for RV segmentation in cardiac MRI. We achieved state-of-the-art performance and presented a thorough comparative study which spans various loss functions, architectures, and ensembling strategies. We establish the effectiveness of the UNet model, a novel switching loss for handling class imbalance, and majority voting ensembling technique for RV segmentation. The thorough comparative study will guide upcoming works in medical image segmentation.

\bibliographystyle{IEEEtran}
\bibliography{bibfile}

% Generated by IEEEtran.bst, version: 1.14 (2015/08/26)
\begin{thebibliography}{10}
\providecommand{\url}[1]{#1}
\csname url@samestyle\endcsname
\providecommand{\newblock}{\relax}
\providecommand{\bibinfo}[2]{#2}
\providecommand{\BIBentrySTDinterwordspacing}{\spaceskip=0pt\relax}
\providecommand{\BIBentryALTinterwordstretchfactor}{4}
\providecommand{\BIBentryALTinterwordspacing}{\spaceskip=\fontdimen2\font plus
\BIBentryALTinterwordstretchfactor\fontdimen3\font minus
  \fontdimen4\font\relax}
\providecommand{\BIBforeignlanguage}[2]{{%
\expandafter\ifx\csname l@#1\endcsname\relax
\typeout{** WARNING: IEEEtran.bst: No hyphenation pattern has been}%
\typeout{** loaded for the language `#1'. Using the pattern for}%
\typeout{** the default language instead.}%
\else
\language=\csname l@#1\endcsname
\fi
#2}}
\providecommand{\BIBdecl}{\relax}
\BIBdecl

\bibitem{petitjean2015right}
C.~Petitjean, M.~A. Zuluaga, W.~Bai, J.-N. Dacher, D.~Grosgeorge, J.~Caudron,
  S.~Ruan, I.~B. Ayed, M.~J. Cardoso, H.-C. Chen \emph{et~al.}, ``Right
  ventricle segmentation from cardiac {MRI}: a collation study,'' \emph{Medical
  image analysis}, vol.~19, no.~1, pp. 187--202, 2015.

\bibitem{guo2018local}
Z.~Guo, W.~Tan, L.~Wang, L.~Xu, X.~Wang, B.~Yang, and Y.~Yao, ``Local motion
  intensity clustering (lmic) model for segmentation of right ventricle in
  cardiac {MRI} images,'' \emph{IEEE journal of biomedical and health
  informatics}, vol.~23, no.~2, pp. 723--730, 2018.

\bibitem{punithakumar2015right}
K.~Punithakumar, M.~Noga, I.~B. Ayed, and P.~Boulanger, ``Right ventricular
  segmentation in cardiac {MRI} with moving mesh correspondences,''
  \emph{Computerized Medical Imaging and Graphics}, vol.~43, pp. 15--25, 2015.

\bibitem{moolan2016right}
O.~Moolan-Feroze, M.~Mirmehdi, and M.~Hamilton, ``Right ventricle segmentation
  using a 3d cylindrical shape model,'' in \emph{2016 IEEE 13th International
  Symposium on Biomedical Imaging (ISBI)}.\hskip 1em plus 0.5em minus
  0.4em\relax IEEE, 2016, pp. 44--48.

\bibitem{ringenberg2014fast}
J.~Ringenberg, M.~Deo, V.~Devabhaktuni, O.~Berenfeld, P.~Boyers, and J.~Gold,
  ``Fast, accurate, and fully automatic segmentation of the right ventricle in
  short-axis cardiac {MRI},'' \emph{Computerized Medical Imaging and Graphics},
  vol.~38, no.~3, pp. 190--201, 2014.

\bibitem{borodin2018right}
G.~Borodin and O.~Senyukova, ``Right ventricle segmentation in cardiac {MR}
  images using u-net with partly dilated convolution,'' in \emph{International
  Conference on Artificial Neural Networks}.\hskip 1em plus 0.5em minus
  0.4em\relax Springer, 2018, pp. 179--185.

\bibitem{winther2018nu}
H.~B. Winther, C.~Hundt, B.~Schmidt, C.~Czerner, J.~Bauersachs, F.~Wacker, and
  J.~Vogel-Claussen, ``$\nu$-net: deep learning for generalized biventricular
  mass and function parameters using multicenter cardiac {MRI} data,''
  \emph{JACC: Cardiovascular Imaging}, p. 2479, 2018.

\bibitem{luo2016deep}
G.~Luo, R.~An, K.~Wang, S.~Dong, and H.~Zhang, ``A deep learning network for
  right ventricle segmentation in short-axis {MRI},'' in \emph{2016 Computing
  in Cardiology Conference (CinC)}.\hskip 1em plus 0.5em minus 0.4em\relax
  IEEE, 2016, pp. 485--488.

\bibitem{tran2016fully}
P.~V. Tran, ``A fully convolutional neural network for cardiac segmentation in
  short-axis {MRI},'' \emph{arXiv preprint arXiv:1604.00494}, 2016.

\bibitem{ronneberger2015u}
O.~Ronneberger, P.~Fischer, and T.~Brox, ``U-net: Convolutional networks for
  biomedical image segmentation,'' in \emph{International Conference on Medical
  image computing and computer-assisted intervention}.\hskip 1em plus 0.5em
  minus 0.4em\relax Springer, 2015, pp. 234--241.

\bibitem{peng2017large}
C.~Peng, X.~Zhang, G.~Yu, G.~Luo, and J.~Sun, ``Large kernel matters--improve
  semantic segmentation by global convolutional network,'' in \emph{Proceedings
  of the IEEE conference on computer vision and pattern recognition}, 2017, pp.
  4353--4361.

\bibitem{github_mri}
C.-H. Yee, ``Cardiac {MRI} segmentation,''
  \url{https://chuckyee.github.io/cardiac-segmentation/}.

\bibitem{petitjean2011review}
C.~Petitjean and J.-N. Dacher, ``A review of segmentation methods in short axis
  cardiac {MR} images,'' \emph{Medical image analysis}, vol.~15, no.~2, pp.
  169--184, 2011.

\bibitem{oghli2018hybrid}
M.~G. Oghli, A.~Mohammadzadeh, R.~Kafieh, and S.~Kermani, ``A hybrid
  graph-based approach for right ventricle segmentation in cardiac {MRI} by
  long axis information transition,'' \emph{Physica Medica}, vol.~54, pp.
  103--116, 2018.

\bibitem{avendi2017automatic}
M.~R. Avendi, A.~Kheradvar, and H.~Jafarkhani, ``Automatic segmentation of the
  right ventricle from cardiac {MRI} using a learning-based approach,''
  \emph{Magnetic resonance in medicine}, vol.~78, no.~6, pp. 2439--2448, 2017.

\bibitem{krahenbuhl2011efficient}
P.~Kr{\"a}henb{\"u}hl and V.~Koltun, ``Efficient inference in fully connected
  crfs with gaussian edge potentials,'' in \emph{Advances in neural information
  processing systems}, 2011, pp. 109--117.

\bibitem{liu2018semi}
X.~Liu, J.~Cao, T.~Fu, Z.~Pan, W.~Hu, K.~Zhang, and J.~Liu, ``Semi-supervised
  automatic segmentation of layer and fluid region in retinal optical coherence
  tomography images using adversarial learning,'' \emph{IEEE Access}, vol.~7,
  pp. 3046--3061, 2018.

\bibitem{lin2017focal}
T.-Y. Lin, P.~Goyal, R.~Girshick, K.~He, and P.~Doll{\'a}r, ``Focal loss for
  dense object detection,'' in \emph{Proceedings of the IEEE international
  conference on computer vision}, 2017, pp. 2980--2988.

\bibitem{paszke2017automatic}
A.~Paszke, S.~Gross, S.~Chintala, G.~Chanan, E.~Yang, Z.~DeVito, Z.~Lin,
  A.~Desmaison, L.~Antiga, and A.~Lerer, ``Automatic differentiation in
  {PyTorch},'' in \emph{NIPS Autodiff Workshop}, 2017.

\bibitem{loshchilov2016sgdr}
I.~Loshchilov and F.~Hutter, ``{SGDR}: Stochastic gradient descent with warm
  restarts,'' \emph{arXiv preprint arXiv:1608.03983}, 2016.

\end{thebibliography}
\end{document}